\documentclass[aps,preprint]{revtex4-1} 
\usepackage{graphicx}
\usepackage{amssymb}
\usepackage{epstopdf}
\usepackage{amsmath}
\usepackage{bm}

\newcommand{\half}{\frac{1}{2}}

\newcommand{\dd}{\partial}
\newcommand{\eps}{\epsilon}

\newcommand\refeq[1]{(\ref{#1})}
\newcommand\reffig[1]{Figure~\ref{#1}}
\newcommand\of[1]{\left( #1 \right)}

\newcommand\vecbf[1]{{\bf #1}}

\begin{document}

\title{Self-Consistent, Self-Coupled Scalar Gravity}

\author{J. Franklin}
\email{jfrankli@reed.edu}

\affiliation{Department of Physics, Reed College, Portland, Oregon 97202,
USA}

\begin{abstract}
A scalar theory of gravity extending Newtonian gravity to include field energy as its source is developed.  The physical implications of the theory are probed through its spherically symmetric (source) solutions.  The aim is to demonstrate rational physical model building, together with physical and experimental checks of correctness.  The theory discussed here was originally considered by Einstein prior to his introduction of general relativity.
\end{abstract}

\maketitle

\section{Introduction}

A field theory describing gravity is necessarily different from other familiar field theories (notably electricity and magnetism) because of universal coupling -- every form of mass/energy acts as a source for gravity, including the energy stored in the gravitational field itself.  That renders gravity theories nonlinear.

General relativity is the modern theory of gravity, developed by Einstein in 1916, and as advertised, the demands of universal coupling lead to nonlinearity in its field equation (Einstein's equation).  Rather than work with the full case, we will explore self-coupling in a scalar field theory -- this simplified setting allows us to think about the definition and role of ``field energy" as a source, and we will end up developing a self-coupled scalar gravity theory that was considered by Einstein in 1912~\cite{Einstein}, re-derived and extended to special relativity in~\cite{FandN}, and re-developed and discussed more recently in~\cite{GiuliniSC}.

The manner in which gravitational field energy acts as a source is not unique -- we'll start by considering a scalar field theory that is self-coupled and in which the field energy density has a particular form~\cite{Peters}.  There is a problem with that approach -- the energy density of a field theory tells us something about its field equations, and the scalar theory we will start with has field equations that are inconsistent with its energy density.

To formalize that problem, and solve it, we will define the field Lagrangian and Hamiltonians by analogy with particle Lagrangians and Hamiltonians.  The field Hamiltonian defines the energy density of a field theory, and that is what must appear in the field equations (derived from an Euler-Lagrange approach for fields) of a ``self-consistent" scalar field theory.  We will arrive at the self-consistent, self-coupled form considered by Einstein, and presented compactly in~\cite{GiuliniSC}.  Once we have the field equations, we will solve them for a few interesting sources, and finally, calculate the precession of the point of closest approach for a body orbiting a spherically symmetric massive source.

The point of the exercise is to take known new physics (the self-energy coupling of the field) and introduce it in the accepted gravitational framework (Newtonian gravity).  That type of targeted, physically-inspired modification, and the evaluation of the physical content of the resulting theory are important elements of modern theoretical model-building.  The scalar case considered here has a familiar starting point in the Poisson equation, with an equally familiar motivation (mass-energy equivalence), and so makes for a good example even though it will not produce the accepted theory of gravity (which has, among other complexities, no unique energy density definition, making it difficult to discuss in the context of field energy-mass equivalence).  Indeed, its failure in that regard is also instructive:  1.\ the theory considered here is scalar, not tensorial, and is only sourced by the $00$ component of the full stress-tensor that must appear in a Lorentz covariant field equation, and 2.\  there is no necessity for a geometric view in the scalar case -- the motivation is not the usual equivalence principles that begin most discussions of general relativity, and are used to establish the geometric interpretation of the theory.

\section{Gravity and Energy}

Newtonian gravity is described by a scalar field $\Phi(\vecbf r)$ that is sourced by mass according to
\begin{equation}\label{Newton}
\nabla^2 \Phi(\vecbf r) = 4 \, \pi \, G \, \rho(\vecbf r)
\end{equation}
where $\rho(\vecbf r)$ is the (specified) mass density.  A particle of mass $m$ moves under the influence of $\Phi$ according to Newton's second law: $m\,  \vecbf a = -m \, \nabla\Phi(\vecbf r)$.  Given a source of mass, you find $\Phi(\vecbf r)$ by solving the above and then you can predict the motion of particles.

\subsection{Energy Source}
In special relativity, we learn that mass and energy are equivalent,  so it is reasonable to consider energetic sources to Newtonian gravity.  As an example, the energy density of the electric field of a point charge with mass $M$ and charge $Q$ sitting at the origin, is:
\begin{equation}
\mathcal{E} = \half \, \epsilon_0 \, \nabla V \cdot \nabla V = \frac{Q^2}{32 \, \pi^2 \, \epsilon_0 \, r^4},
\end{equation}
where $V$ is the electrostatic potential associated with a point charge.  We could use this energy density as a source in~\refeq{Newton} by noting that the associated mass density is $\rho_{\mathcal{E}} = \mathcal{E}/c^2$, and then we need to solve
\begin{equation}\label{NewtonCharge}
\nabla^2 \Phi(\vecbf r) =4 \, \pi \, G\,  \frac{1}{c^2} \, \left[ \frac{Q^2}{32 \, \pi^2 \, \epsilon_0 \, r^4} \right]
\end{equation}
for $\Phi$, keeping in mind that there is also a point mass source at the origin.  The solution is
\begin{equation}\label{ChargePP}
\Phi(\vecbf r) = -\frac{G \, M}{r} + \frac{G \, Q^2}{16 \, \pi \, \eps_0 \, c^2\, r^2}.
\end{equation}
This gravitational field contains a charge contribution, meaning, among other things, that a neutral massive body moving under the influence of this potential would be able to detect the presence of the central charge.  This result agrees in form and interpretation with the linearized result from general relativity (where we find an effective potential of the form~\refeq{ChargePP} from the metric associated with the exterior of a spherically symmetric charged central body, see~\cite{MTW}).

\subsection{Self-Source}

The gravitational field, like the electric one, carries energy, and so we might start by asking how $\Phi$ sources itself.  The first answer comes by analogy with E\&M (see~\cite{Griffiths} for a review of that process).  The work required to build a configuration of mass with mass density $\rho(\vecbf r)$ and associated potential $\Phi(\vecbf r)$ is given by
\begin{equation}\label{workdone}
W = \half \, \int_{\hbox{\tiny{all space}}}  \rho(\vecbf r) \, \Phi(\vecbf r) \, d\tau.
\end{equation}
Using~\refeq{Newton}, we can replace $\rho(\vecbf r)$ with $\frac{1}{4 \, \pi \, G} \, \nabla^2 \Phi(\vecbf r)$ and integrate by parts, assuming that $\Phi(\vecbf r)$ vanishes at spatial infinity (the boundary of our integration domain) to get
\begin{equation}\label{energystored}
W = -\int_{\hbox{\tiny{all space}}}  \frac{1}{8 \, \pi \, G} \, \nabla \Phi(\vecbf r)\cdot\nabla\Phi(\vecbf r) \, d\tau.
\end{equation}
The integrand is an energy density that we interpret as the energy density associated with $\Phi(\vecbf r)$: $\mathcal{E} = -\frac{1}{8 \, \pi \, G} \, \nabla \Phi(\vecbf r)\cdot\nabla\Phi(\vecbf r)$.  So it is reasonable to
start with
\begin{equation}\label{PetersP}
\nabla^2 \Phi(\vecbf r) = 4 \, \pi \, G \, \rho(\vecbf r) - \frac{1}{2 \, c^2} \, \nabla \Phi \cdot \nabla \Phi,
\end{equation}
precisely as is developed in~\cite{Peters}.

The argument here mimics the usual one from E\&M, but there is a crucial difference -- we end up modifying the Poisson form of the field equation: ~\refeq{Newton} becomes~\refeq{PetersP}.  But in writing the energy density (i.e.\ in going from~\refeq{workdone} to~\refeq{energystored}), we {\it used}~\refeq{Newton}, while we should have used~\refeq{PetersP}.  The self-sourcing in this case is, in a sense, inconsistent with the field equation.

\section{Classical Field Theory}

Here we'll introduce just enough field theory to sensibly define the energy density of a scalar field.  We'll work by analogy with the Lagrangian and Hamiltonian from particle mechanics, and start with a brief review.

\subsection{Particles}

The Lagrangian for a particle reads: $L = \half \, m \, \dot{\vecbf r}(t) \cdot \dot{\vecbf r}(t) - U(\vecbf r(t))$ where $U$ is potential energy and the first term represents the kinetic energy (for position vector $\vecbf r(t)$, we define the velocity vector: $\dot{\vecbf r}(t) \equiv \frac{d \vecbf r(t)}{d t}$).  The Euler-Lagrange equations of motion for a particle moving under the influence of $U$ are:
\begin{equation}\label{EL}
\frac{d}{dt} \, \of{ \frac{\dd L}{\dd \dot{\vecbf r}} } - \frac{\dd L}{\dd \vecbf r} = 0,
\end{equation}
a total of $D$ equations in $D$ dimensions.  Related to the Lagrangian is the Hamiltonian: $H = \frac{1}{2 \, m} \, \vecbf p(t) \cdot \vecbf p(t) + U(\vecbf r(t))$ (for momentum $\vecbf p(t) \equiv \frac{\dd L}{\dd \dot{\vecbf r}}$), representing the total energy of the system.   For a ``free particle", there is no potential, and $L = H$, the Lagrangian is numerically identical to the Hamiltonian, and we can write $H = \half \, m \, \dot{\vecbf r}(t)\cdot \dot{\vecbf r}(t)$.

\subsection{Fields}

Time-independent fields, like $\Phi(\vecbf r)$, have ``Lagrange densities" (for a review of this type of approach to classical field theory, see~\cite{Franklin}, for example) given by ${\mathcal L} = \half \, \alpha \, \nabla \Phi \cdot \nabla \Phi - W(\Phi)$ where $\alpha$ is a constant that sets the dimension of the Lagrange density, and $W(\Phi)$ is a field ``potential".  The ``equations of motion", called field equations here, are given by
\begin{equation}\label{ELF}
\nabla \cdot \of{\frac{\dd {\mathcal{L}}}{\dd \nabla \Phi}}  - \frac{\dd {\mathcal{L}}}{\dd \Phi} = 0,
\end{equation}
which is just like~\refeq{EL}, but with time-derivatives replaced by spatial ones: $\frac{\dd L}{\dd \of{d\vecbf r/dt}} \longrightarrow \frac{\dd \mathcal{L}}{\dd\of{\nabla \Phi}}$.  As an example, take 
\begin{equation}\label{NL}
\mathcal{L} =\frac{1}{4 \, \pi \, G} \, \half \,  \nabla \Phi \cdot \nabla \Phi - (- \rho \, \Phi)
\end{equation}
for a specified mass density $\rho$ ({\it not} a function of $\Phi$).  In the first term, we have set $\alpha \equiv \frac{1}{4 \, \pi \, G}$ so that the Lagrange density has dimension of energy-over-volume (an energy density).  The second term, representing $W(\Phi)$ is linear in the field, typical of ``source" terms.

For the field equation, we have $\of{\frac{\dd {\mathcal{L}}}{\dd \nabla \Phi}}  =\frac{1}{4 \, \pi \, G} \, \nabla \Phi$, so that $\nabla \cdot \of{\frac{\dd {\mathcal{L}}}{\dd \nabla \Phi}}  = \frac{1}{4 \, \pi \, G} \, \nabla^2 \Phi$.  The other term is $ \frac{\dd {\mathcal{L}}}{\dd \Phi} = \rho$, and we can put these pieces together in~\refeq{ELF} to get precisely the field equation for Newtonian gravity:
\begin{equation}\label{NewtonfromL}
\frac{1}{4 \, \pi \, G} \, \nabla^2\Phi - \rho = 0.
\end{equation}

There is also a Hamiltonian density associated with $\Phi$: ${\mathcal H} = \half \, \alpha \, \nabla \Phi \cdot \nabla \Phi + W(\Phi)$, and again, in the absence of $W$ (for a ``free" field), we have $\mathcal{L} = \mathcal{H}$, and this is the energy density of the field itself.  For the (free, $\rho = 0$) Lagrange density associated with gravity, we would call the energy density $\mathcal{L} = \mathcal{H} = \frac{1}{8\, \pi\, G} \, \nabla \Phi \cdot \nabla \Phi$ from~\refeq{NL}, and this agrees with the field energy density we got in~\refeq{energystored}.  But this density does not itself appear as a source in~\refeq{NewtonfromL}, and if we insert it by hand, the resulting field equation does not come from the starting Lagrangian~\refeq{NL} (with $\rho = 0$).   There is, then, a clash between the energy density defined by the Lagrangian~\refeq{NL} and the field equations that come from that Lagrangian, which do not take the form~\refeq{PetersP}.  What we want to do is develop a Lagrangian (and hence an energy density expression) whose field equation, coming from the Lagrangian, involves that energy density.  Then we will have a ``self-consistent" theory.

\section{Self-Consistent, Self-Sourced Gravity}

In this section, we will engineer a field equation that has its own energy density as a source.  We'll work in vacuum, and then add in massive sources $\rho(\vecbf r)$ at the end.   Our target is a Lagrange density $\mathcal{L}$ such that the field equation (again, in the absence of explicit sources) is $\nabla^2 \Phi = 4 \, \pi \, G \, \frac{\mathcal{L}}{c^2}$ ($\mathcal{L}$ is the free-field energy density, acting here as a source).  The field equation will be self-consistent if it comes from~\refeq{ELF} for our chosen $\mathcal{L}$.

Start with 
\begin{equation}
\mathcal{L} = \frac{f(\Phi)}{8 \, \pi \, G} \, \nabla\Phi \cdot \nabla \Phi,
\end{equation}
for an arbitrary dimensionless function of $\Phi$, $f(\Phi)$.  The plan is to get the field equation for this $\mathcal{L}$ from~\refeq{ELF}, and then fix $f(\Phi)$ by requiring that the right hand side of that field equation be $4 \, \pi \, G \, \frac{\mathcal{L}}{c^2}$.  We have: 
\begin{equation}
\frac{\dd {\mathcal L}}{\dd \nabla \Phi} = \frac{f(\Phi)}{4 \, \pi \, G} \,\nabla\Phi \longrightarrow \nabla \cdot \frac{\dd {\mathcal L}}{\dd \nabla \Phi} = \frac{f'(\Phi)}{4 \, \pi \, G} \, \nabla \Phi \cdot \nabla \Phi + \frac{f(\Phi)}{4 \, \pi  \, G} \, \nabla^2 \Phi.
\end{equation}
The other term in~\refeq{ELF} is: $\frac{\dd {\mathcal L}}{\dd \Phi} = \frac{f'(\Phi)}{8 \, \pi \, G} \, \nabla\Phi \cdot \nabla\Phi$.  Putting the pieces together, the vacuum field equation is
\begin{equation}
\frac{f(\Phi)}{4 \, \pi \, G} \, \nabla^2 \Phi  +  \frac{f'(\Phi)}{8 \, \pi \, G} \, \nabla\Phi \cdot \nabla\Phi = 0 \longrightarrow \nabla^2 \Phi = -\frac{f'(\Phi)}{2 \, f(\Phi)} \, \nabla \Phi \cdot \nabla\Phi.
\end{equation}
Looking at the right-hand side, we would have an energy density source if:
\begin{equation}
 -\frac{f'(\Phi)}{2 \, f(\Phi)} \, \nabla \Phi \cdot \nabla\Phi = 4 \, \pi \, G \, \frac{\mathcal L}{c^2} = \frac{f(\Phi)}{2 \, c^2} \, \nabla \Phi \cdot \nabla \Phi,
 \end{equation}
or $-\frac{f'(\Phi)}{f(\Phi)} = \frac{f(\Phi)}{c^2}$.  This can be solved for $f(\Phi)$, defining our original Lagrangian: $f(\Phi) = \frac{c^2}{\Phi}$.

The starting Lagrangian is now $\mathcal L = \frac{c^2}{8 \, \pi \, G \, \Phi} \, \nabla\Phi \cdot \nabla\Phi$.  If we add back in the mass density source $\rho(\vecbf r)$, we get
\begin{equation}
\mathcal{L} =  \frac{c^2}{8 \, \pi \, G \, \Phi} \, \nabla\Phi \cdot \nabla\Phi + \rho\, \Phi.
\end{equation}
The associated field equations, from~\refeq{ELF}, are
\begin{equation}\label{SCSC}
\nabla^2 \Phi = \frac{4 \, \pi \, G}{c^2} \, \rho \, \Phi + \frac{1}{2 \, \Phi} \, \nabla \Phi \cdot \nabla\Phi.
\end{equation}
This is the field equation obtained in~\cite{GiuliniSC} and developed earlier, {\it en route} to general relativity in~\cite{Einstein}.  It represents a scalar (meaning here that $\Phi$ is just a number, not a vector or tensor) theory of gravity with self-consistent self-sourcing.  If we had performed the analogous procedure using a symmetric second rank tensor field (which could be represented as a symmetric matrix), we would have obtained General Relativity as in~\cite{Deser}.  This program was carried out both in a scalar self-interaction (to the trace of its stress-tensor) and in the geometric (a la GR) settings -- restricted to a conformally flat metric -- in~\cite{DandH}.

Before moving on, it is worth noting that while special relativity motivated the introduction of the self-sourcing, the resulting static field equations in~\refeq{SCSC} are clearly not Lorentz covariant, lacking the temporal portion that would need to be in place to maintain scalar character under Lorentz transformation \footnote{We could accomplish this by taking $\nabla \rightarrow \dd_\mu$, and $\nabla^2 \rightarrow \dd_\mu \, \dd^\mu \equiv \Box$, and study the resulting dynamics.}.  Assuming we could correct that problem, we would still have a non-scalar element in~\refeq{SCSC}, namely $\rho$.  A mass density like $\rho$ is {\it not} a scalar under Lorentz transformation (owing in part to the volume dependence of the mass {\it density}), and in fact one can show (by transformation) that $\rho$ is the $00$ component of a second-rank tensor, meaning that~\refeq{SCSC}, even with appropriate temporal dependence, would lack the Lorentz covariance required of a theory of gravity wholly in accord with special relativity.

Those considerations aside (and they can be used to continue modifying and correcting until all of general relativity is obtained), we can write~\refeq{SCSC} linearly in $\sqrt{\Phi}$, following~\cite{Einstein}:
\begin{equation}\label{SCSCE}
\nabla^2 \of{\sqrt{\Phi}} = \frac{2 \pi \, G}{c^2} \, \rho \, \sqrt{\Phi}
\end{equation}
which is easier to solve.  We'll explore the spherically symmetric solutions next.

\section{Solutions}

\subsection{Spherically symmetric vacuum}
We'll start by looking at solutions to the field equation~\refeq{SCSCE} where $\rho = 0$ (so we are in vacuum) and $\Phi(\vecbf r) = \Phi(r)$.  Such a solution would be appropriate outside of a spherically symmetric source of mass $M$ localized near the origin.  Under our assumptions, the field equation~\refeq{SCSCE} becomes the Laplace equation for $\sqrt{\Phi(r)}$ and that is easily solved:
\begin{equation}
\nabla^2 \left(\sqrt{\Phi(r)}\right) = \frac{1}{r} \, \left( r \, \sqrt{\Phi(r)} \right)'' = 0 \longrightarrow \sqrt{\Phi(r)} = \frac{\alpha}{r} + \beta.
\end{equation}
with primes denoting $r$-derivatives.  We can write $\Phi(r)$ in terms of the two integration constants, $\alpha$ and $\beta$,
\begin{equation}
\Phi(r) =\beta^2 +  \frac{2 \, \alpha \, \beta}{r} + \frac{\alpha^2}{r^2}.
\end{equation}
If we ask that $\Phi(r)$ look like the Newtonian potential of a point source (of mass $M$) as $r$ approaches spatial infinity, then we can fix $\alpha$
\begin{equation}
\Phi(r) =\beta^2  -\frac{G \, M}{r} + \frac{G^2 \, M^2}{4 \, \beta^2 \, r^2}.
\end{equation}
Now we must choose $\beta$ -- it cannot be zero, and it must have a value with the dimensions of speed, so that $c$ is a natural candidate.  The term $\beta$ is dominant far from the source, in the weak field regime.  Going back to the field equation~\refeq{SCSC}, if we set $\Phi = c^2 + \Phi_N$ there, and collect in powers of $c$, the leading PDE is precisely Newtonian gravity for $\Phi_N$: $\nabla^2 \Phi_N = 4 \, \pi \, G \, \rho$.  So in the weak field regime, we expect $\Phi = c^2 + \Phi_N$ -- that means we {\it must} set $\beta = c$ in our vacuum solution.

Our final spherically symmetric vacuum solution looks like
\begin{equation}\label{ppSC}
\Phi(r) = -\frac{G \, M}{r} + \frac{G^2 \, M^2}{4 \, c^2 \, r^2} + c^2.
\end{equation}
The term that goes like $1/r^2$ is similar to the one appearing in~\refeq{ChargePP}.  Indeed, if we take the na\"ive replacement $\frac{1}{4 \, \pi \, \epsilon_0} \rightarrow G$ that turns the magnitude of electrostatic results into Newtonian gravitational ones, we would expect the charge term in~\refeq{ChargePP} to turn into
\begin{equation}
\frac{G\, Q^2}{16 \, \pi \, \epsilon_0 \, c^2 \, r^2} \longrightarrow \frac{G^2 \, M^2}{4 \, c^2 \, r^2},
\end{equation}
and this is precisely the term we find in~\refeq{ppSC}.

\subsection{Spherically symmetric, constant $\rho$}

If we take a sphere of radius $R$ with $\rho(\vecbf r) = \rho_0$, a constant, we again expect spherically symmetric $\Phi$, and~\refeq{SCSCE} becomes:
\begin{equation}
\frac{1}{r} \, \left( r \, \sqrt{\Phi}\right)'' = \frac{2 \, \pi\, G \,\rho_0 }{c^2} \, \sqrt{\Phi} \longrightarrow  \left( r \, \sqrt{\Phi}\right)''  = \frac{1}{r_0^2} \, \left( r \, \sqrt{\Phi} \right),
\end{equation}
with $r_0 \equiv \frac{c}{\sqrt{2 \, \pi \, G \, \rho_0}}$.  In this form, it is easy to see that the solution for $r \, \sqrt{\Phi}$ is just a sum of sinh and cosh:
\begin{equation}
r \, \sqrt{\Phi} = A \, \sinh\of{\frac{r}{r_0}} + B \, \cosh\of{\frac{r}{r_0}}.
\end{equation}
We want a solution for $\Phi$ that is finite at the origin, so set $B = 0$ and keep the sinh term.  Then the interior solution is:
\begin{equation}
\Phi_i(r) = \left(\frac{A \, \sinh\of{r/r_0} }{r}\right)^2.
\end{equation}

\begin{figure}[htbp] %  figure placement: here, top, bottom, or page
   \centering
   \includegraphics[width=4in]{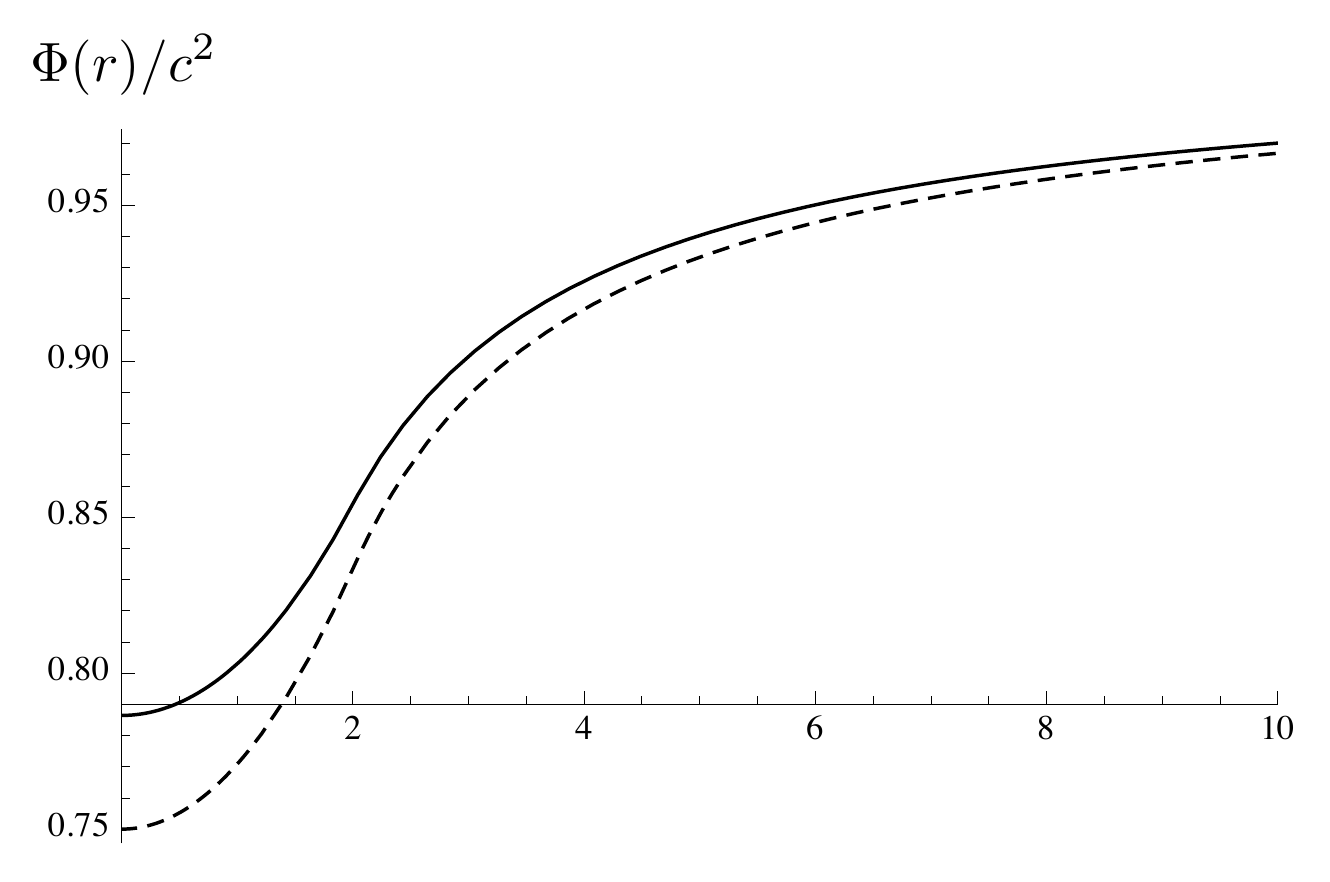} 
   \caption{Self-coupled (solid) and Newtonian (dashed) potential for a homogenous solid sphere of mass (with $R=2$, $r_0 = 4$).}
   \label{fig:comp}
\end{figure}

Outside the sphere of mass, where $\rho = 0$, the exterior potential is as above:
\begin{equation}
\Phi_o(r) = \left(\frac{\alpha}{r} + \beta\right)^2.
\end{equation}
If we demand both continuity and derivative-continuity at the boundary of the central sphere, so that $\Phi_i(R) = \Phi_o(R)$ and $\Phi'_i(R) = \Phi'_o(R)$, and in addition, we set $\Phi_o(\infty) = c^2$, then we can fix all remaining constants, to get
\begin{equation}
\begin{aligned}
\Phi_i(r) &= \of{ \frac{c}{\cosh\of{\frac{R}{r_0}}} \, \frac{\sinh\of{\frac{r}{r_0}}}{\frac{r}{r_0}} }^2 \\
\Phi_o(r) &= c^2 \, \left[ -\frac{2\, \bar R}{r} + \frac{\bar R^2}{r^2} + 1 \right] \\
\bar R&\equiv  \of{R - r_0 \, \tanh\of{\frac{R}{r_0}} }.
\end{aligned}
\end{equation}
Compare with the Newtonian result (for the same density and boundary conditions):
\begin{equation}
\begin{aligned}
\Phi_i(r) &= \frac{c^2 \, r^2}{3 \, r_0^2} - \frac{c^2 \, \of{R^2 - r_0^2}}{r_0^2} \\
\Phi_o(r) &= c^2 \, \left[ -\frac{2 \of{\frac{R^3}{3 \, r_0^2}}}{r}  + 1 \right].
\end{aligned}
\end{equation}
In~\reffig{fig:comp}, we can see the self-coupled (solid curve) and Newtonian (dashed curve) potentials for a sphere with $R = 2$ and $r_0 = 4$.

\subsection{Charged, massive central body}

We can also calculate the effect of charge $Q$ placed on a point mass $M$ in the self-coupled case (as we did earlier for Newtonian gravity, where the result is~\refeq{ChargePP}, and it was done for the field equation~\refeq{PetersP} in~\cite{Young}).  If we use: $\rho = \frac{Q^2}{32 \, \pi^2 \, \eps_0 \, r^4 \, c^2}$ as in~\refeq{NewtonCharge}, 
and again assume spherical symmetry, then we can write~\refeq{SCSCE} as
\begin{equation}\label{cstart}
\frac{1}{r} \, \frac{d^2}{dr^2} \, \left( r \, \sqrt{\Phi}\right) = \frac{Q^2 \, G}{16 \, \pi \, \epsilon_0} \, \frac{1}{c^4} \, \frac{\sqrt{\Phi}}{r^4}.
\end{equation}
Define $\tilde r^2 \equiv \frac{Q^2 \, G}{16 \, \pi \, \epsilon_0\, c^4}$, and note that for $p \equiv 1/r$, we have:
\begin{equation}
\frac{d^2}{dr^2} \, \left( r \, \sqrt{\Phi} \right) = \frac{1}{r^3} \, \frac{d^2}{d p^2} \, \sqrt{\Phi},
\end{equation}
then we can simplify and solve~\refeq{cstart}:
\begin{equation}
\frac{d^2}{d p^2} \, \sqrt{\Phi} = \tilde r^2 \, \sqrt{\Phi} \longrightarrow \sqrt{\Phi} = \alpha \, \cosh\of{\tilde r/r} + \beta \, \sinh\of{\tilde r/r}.
\end{equation}
From this, we can expand near spatial infinity to match the constant and $1/r$ terms to the asymptotic Newtonian form: $c^2 -\frac{G \, M}{r}$.  As $r \rightarrow \infty$
\begin{equation}
\Phi \stackrel{r \rightarrow \infty}{\longrightarrow} \alpha^2 + \frac{2 \, \alpha\, \beta \, \tilde r}{r} + O\left( \left(\frac{1}{r}\right)^2 \right)
\end{equation}
so set $\alpha = c$ and $\beta = -\frac{G \, M}{2 \, c \, \tilde r}$ giving
\begin{equation}
\Phi(r) = \left( c \, \cosh(\tilde r/r) - \frac{G \, M}{2 \, c \, \tilde r} \sinh(\tilde r/r) \right)^2.
\end{equation}
It is interesting to note that if we expand this at spatial infinity, and display more terms, we get:
\begin{equation}
\Phi(r) \stackrel{r \rightarrow \infty}{\longrightarrow} c^2 - \frac{G \, M}{r} + \frac{G^2 \, M^2}{4 \, c^2 \, r^2} + \frac{G \, Q^2}{16 \, \pi \, \epsilon_0\, c^2 \, r^2} + O\left( \left(\frac{1}{r}\right)^3 \right)
\end{equation}
where the first three terms match the self-coupled vacuum potential from~\refeq{ppSC} and we recognize the first charge term from~\refeq{ChargePP}.
\section{Precession}

Turning to the other element of this new gravitational field theory -- the equations of motion -- we can calculate the dynamics of a massive body moving under the influence of any of the above potentials.  
Bertrand's theorem (see, for example~\cite{Goldstein}) tells us that for any spherically symmetric potential other than $\sim r^2$ and $\sim 1/r$, elliptical orbits will precess, and all of the spherically symmetric potentials developed above have terms other than $1/r$.  We'll take the vacuum solution~\refeq{ppSC} for simplicity, and calculate the precession for a body of mass $m$ going around a spherical central body of mass $M$.  Starting from the Hamiltonian:
\begin{equation}
H =E = \half \, m \, \of{ \dot r^2 + \frac{J_z^2}{r^2} } + U(r) \, \, \, \, \, \, \, \, \, \, \, \dot\phi = \frac{J_z}{m \, r^2}
\end{equation}
for a particle of mass $m$ with angular momentum ($z$-component) $J_z$, if we take $p = 1/r$ and parametrize in terms of $\phi$ rather than $t$, we get:
\begin{equation}
p'(\phi)^2 = \frac{2 \, m}{J_z^2} \, \of{E - U(p)} - p(\phi)^2.
\end{equation}
Taking the $\phi$-derivative gives us an equation of motion for $p$:
\begin{equation}
p''(\phi) = -\frac{m}{J_z^2} \, U'(p) - p(\phi),
\end{equation}
and we can use the potential energy:  $U(p) = m \, \of{ -G \, M\, p + \frac{G^2 \, M^2}{4 \, c^2} \, p^2 + c^2}$ to get
\begin{equation}
p''(\phi)  = \frac{G \, M \, m^2}{J_z^2} - p(\phi) \, \of{1 + \frac{G^2 \, M^2\, m^2}{2 \, J_z^2 \, c^2} }.
\end{equation}
The solution here can be written in terms of two arbitrary constants, and we can specialize to the orbital case, where we have:
\begin{equation}
p(\phi) = A + B \, \cos\of{\sqrt{1 + \frac{G^2 \, M^2\, m^2}{2 \, J_z^2 \, c^2}}  \, \phi}
\end{equation}
with $A \equiv \frac{2 \, M \, c^2 \, G \, m^2}{2 \, c^2 \, J_z^2 + G^2 \, m^2 \, M^2}$, and $B$ is a constant of integration.  Now the radial coordinate is just the inverse of $p$, so
\begin{equation}
r(\phi) = \frac{1}{A + B \, \cos\of{\sqrt{1 + \frac{G^2 \, M^2\, m}{2 \, J_z^2 \, c^2}}  \, \phi}}.
\end{equation}
The point of closest approach to the central body occurs at $\phi = 0$, and then again at $\phi_1$ given by:
\begin{equation}
\sqrt{1 + \frac{G^2 \, M^2\, m}{2 \, J_z^2 \, c^2}}  \, \phi_1 = 2 \, \pi \longrightarrow \phi_1 = \frac{2 \, \pi}{\sqrt{1 + \frac{G^2 \, M^2\, m}{2 \, J_z^2 \, c^2}} }.
\end{equation}
If we take  $\frac{G^2 \, M^2\, m^2}{2 \, J_z^2 \, c^2} \ll 1$, so a small departure from elliptical, then the orbiting particle returns to the closest approach point not at $\phi_1 = 2 \, \pi$, but rather at
\begin{equation}
\phi_1 \approx 2\, \pi \, \of{ 1- \frac{G^2 \, M^2\, m^2}{4 \, J_z^2 \, c^2}} < 2 \, \pi.
\end{equation}
The closest approach point then shifts, at each period, to a smaller angle -- this is the opposite of the observed direction for Mercury (for example, which has the analogous $\phi_1 > 2 \, \pi$).

\section{Conclusion}

We have explored the role of self-sourcing in a scalar theory of gravity.  That self-sourcing, when done in a self-consistent manner leads to a new (as compared to Newtonian gravity) gravitational field equation, with new solutions for spherically symmetric sources.  Those solutions, while sensible, predict different physics than Newtonian gravity.  In itself, that is not a problem, since general relativity does as well.  Exploring the physical implications of any modification to a known theoretical framework is as important as generating the modification, and here we could easily compute the observable closest approach precession due to a spherically symmetric massive source.  That precession goes in the opposite direction of both the one predicted by general relativity, and the one observed in, for example, the orbit of Mercury (this is also true in Nordstr\"om's scalar theory of gravity).

Regardless, the utility of the current work is not in extending gravity, but rather, in displaying a logical extension to Newtonian gravity -- one that is physically motivated, and relevant to the ``derivation" of general relativity itself -- when the techniques used in the current work are applied in the proper setting, general relativity emerges, together with its geometric interpretation, quite naturally.    

\acknowledgements
The author thanks David Griffiths for useful commentary and physical insight, and Stanley Deser for additional scalar gravity elements.

%
%\vskip .5in
%\noindent Figure 1. Self-coupled (solid) and Newtonian (dashed) potential for a homogenous solid sphere of mass (with $R=2$, $r_0 = 4$).

\end{document}